\documentclass[11pt,a4paper]{article}
\usepackage{jheppub}
\usepackage[T1]{fontenc}
\usepackage{amsmath}
\usepackage{amssymb}
\usepackage{esint}
\usepackage{amsfonts}
\usepackage{graphicx,color}
\usepackage{slashed}
\usepackage{young}
\usepackage[utf8]{inputenc}
\usepackage{etoolbox}
\usepackage{bbm}
\usepackage{soul}

\makeatletter

\newcommand{\be}{\begin{equation}}
\newcommand{\ee}{\end{equation}}
\newcommand{\bea}{\begin{eqnarray}}
\newcommand{\eea}{\end{eqnarray}}

\renewcommand{\d}{\partial}

\def\cL{\mathcal{L}}

\newcommand*\xbar[1]{%
  \hbox{%
    \vbox{%
      \hrule height 0.5pt 
      \kern0.3ex
      \hbox{%
        \kern-0.0em
        \ensuremath{#1}%
        \kern-0.0em
      }%
    }%
  }%
} 

\pdfpageheight\paperheight
\pdfpagewidth\paperwidth

\pdfoutput=1 

    \patchcmd{\maketitle}{\@fpheader}{}{}{}

\title{\boldmath A note on electric-magnetic duality and soft charges}

\author[a,b,1]{Marc Henneaux\note{\href{https://orcid.org/0000-0002-3558-9025}{ORCID: 0000-0002-3558-9025}},}
\author[c]{Cédric Troessaert}

\affiliation[a]{Universit\'e Libre de Bruxelles and International Solvay Institutes,\\
  Physique Math\'ematique des Interactions Fondamentales,\\
  Campus Plaine---CP~231,
  Bruxelles B-1050, Belgium}
\affiliation[b]{Collège de France, 11 place Marcelin Berthelot, 75005
  Paris, France}
\affiliation[c]{Haute-Ecole Robert Schuman, Rue Fontaine aux M\^ures, 13b, B-6800, Belgium }

\emailAdd{henneaux@ulb.ac.be}
\emailAdd{cedric.troessaert@hers.be}

\abstract{We derive the asymptotic symmetries of the manifestly duality invariant formulation of electromagnetism in Minkoswki space.  We show that the action is invariant under  two algebras of angle-dependent $u(1)$ transformations, one electric and the other magnetic.  As in the standard electric formulation, Lorentz invariance requires the addition of additional boundary degrees of freedom at infinity, found here to be of both electric and magnetic types. A notable feature of this duality symmetric formulation, which we comment upon,  is that the on-shell values of the zero modes of the gauge generators are equal to only half of the electric and magnetic fluxes (the other half is brought in by Dirac-string type contributions).  Another notable feature is the absence of central extension in the angle-dependent $u(1)^2$-algebra.}

\makeatother

\begin{document} 
\maketitle
\flushbottom

\section{Introduction}
\label{sec:introduction}

The asymptotic structure of electromagnetism in Minkowski space has
been a subject of great interest in the last years, with the discovery
that soft photon theorems could be viewed as Ward identities of the
corresponding asymptotic symmetries \cite{He:2014cra}, triggering a
lot of insightful activity
\cite{Lysov:2014csa,Kapec:2014zla,Kapec:2015ena,Campiglia:2016hvg,Conde:2016csj}
reviewed in \cite{Strominger:2017zoo}. (Earlier work on the asymptotic symmetries of electromagnetism at null infinity involves
\cite{Strominger:2013lka,Barnich:2013sxa}.)

While this work was originally focused on null infinity, the structure
of the asymptotic symmetry algebra, which is given by arbitrary
functions on the $2$-sphere (``angle-dependent $u(1)$
transformations'') was also explored at spatial infinity
\cite{Balachandran:2013wsa,Campiglia:2017mua,Henneaux:2018gfi} and equivalence between the
two formulations demonstrated. In particular the antipodal matching
conditions of the null infinity approaches, relating fields at the
past of $\mathcal{I}^+$  to fields at the future of $\mathcal{I}^-$ could be
justified on a dynamical basis \cite{Henneaux:2018gfi}. The proof of
equivalence involves an interesting change of basis in the algebra
based on a parity decomposition.

The above formulations are ``purely electric'' and exhibit only one angle-dependent $u(1)$ symmetry.  As shown in \cite{Strominger:2015bla}, there exists a second angle-dependent $u(1)$ symmetry.  It corresponds to ``large'' gauge transformations acting on the dual potential and can be exibited in the magnetic formulation.
  
There exists a formulation of electromagnetism in which electric-magnetic duality, which is always a symmetry of the action \cite{Deser:1976iy}, is manifest. This formulation is first-order and involves two vector potentials, which are not only duality-conjugate, but also canonically conjugate.  Duality-symmetry is then a bona fide Noether symmetry of standard type \cite{Deser:1976iy}. This symmetry extends to a $sp(n)$ symmetry of the action - and not just of the field equations - when scalar field couplings of appropriate form are included \cite{Bunster:2011aw}.  Although the formulation of  \cite{Deser:1976iy} involves two vector potentials, we stress that it is equivalent to the standard one-potential formulation of Maxwell theory.  The equations of motion for the two vector potentials are of first order, so that the amount of physical, free initial data is unchanged. In particular, there is only one photon, and only two physical degrees of freedom per space point.

In the manifestly duality-invariant formulation, each vector potential enjoys a separate $u(1)$ gauge symmetry.  The gauge transformations can be either ``proper'' or ``improper'' \cite{Benguria:1976in}, depending on their behaviour at infinity. The purpose of this note is to show that the two angle-dependent $u(1)$ symmetries separately displayed  in either the pure electric or the pure magnetic formulations, are actually both simultaneously present as standard improper gauge symmetries  in the duality-invariant formulation.  None of these groups of transformations needs a special treatment and both follow from the application of standard rules. Their generators, computed through canonical methods, are given by non-vanishing surface integrals, which can easily be written down in terms of the variables of the duality-symmetric formulation.

As explained in \cite{Deser:1997mz}, the coupling to sources involves both minimal coupling terms and Dirac-type coupling terms with Dirac strings \cite{Dirac:1948um}.  Each type of couplings contributes half of the total coupling in the duality-symmetric formulation.  It follows that the generators of improper gauge transformations, associated with minimal couplings, only gives half of the total electric or magnetic fluxes.  This factor of one-half is not the result of an incorrect symplectic structure but is built in the construction.  Also built in the approach of \cite{Deser:1997mz} is the fact that the electric and magnetic charges are non dynamical $c$-numbers. They have therefore zero Poisson bracket with any quantity.  This precludes the appearance of a central charge in the algebra of the angle-dependent $u(1)$ generators, found in other, different treatments \cite{Hosseinzadeh:2018dkh,Freidel:2018fsk}.

Our paper is organized as follows.  In Section \ref{sec:Starting}, we briefly recall the duality invariant formulation of \cite{Deser:1976iy,Deser:1997mz} (see also \cite{Henneaux:1988gg,Schwarz:1993vs}) and discuss the specific features due to the degeneracy of the pre-symplectic form that follows from the action.  We give boundary conditions on the vector potentials, which involve parity conditions with a twist given by a gradient, extending the work of \cite{Henneaux:2018gfi} to the double-potential formulation.  We then derive (some of) the improper gauge transformations.  Section \ref{sec:Poincare} turns to Poincar\'e invariance.  We introduce surface degrees of freedom at infinity along the lines of our previous treatment \cite{Henneaux:2018gfi}, which are necessary to make the boosts preserve the pre-symplectic structure.  These surface degrees of freedom, somewhat reminiscent of those introduced in \cite{Gervais:1976ec}, bring in their own improper gauge transformations, which are written.  All the improper gauge transformations combine to form an angle-dependent $u(1)^2$ symmetry, which is the same as the one found in null infinity analyses. The Poisson bracket algebra of all the improper gauge transformations is shown not to involve a central extension.  The $so(2)$ duality rotation is also worked out and contains, in addition to the Chern-Simons term found in  \cite{Deser:1976iy}, an extra contribution involving the new surface degrees of freedom.  Section \ref{sec:Sources} briefly comments on the introduction of sources and the need to introduce parity-symmetric Dirac strings, as in \cite{Bunster:2006rt}.  Finally, we close our paper with comments on the dressings of physical states (Section \ref{sec:Conclusions}).

\section{Starting point}
\label{sec:Starting}

\subsection{Action and presymplectic form}
\label{sec:action}

We start with the first-order manifestly duality invariant action
\cite{Deser:1976iy,Deser:1997mz},
\begin{equation}
  \label{eq:action}
S[A^a] = \int dt \int d^3x \left(\frac12 \epsilon_{ab} \epsilon^{ijk} \partial_i A^a_j \dot{A}^b_k - \mathcal{H} \right)
\end{equation}
with\footnote{Latin indices from the beginning of the alphabet ($a, b, \ldots$) are ``internal'' indices taking the values
$1$ and $2$.  The internal Levi-Civita epsilon tensor is $\epsilon_{ab}$, with
$\epsilon_{12}=1$. Latin indices from the middle of the alphabet ($i, j, k, \ldots$) run
from $1$ to $3$ ($\epsilon_{123}=1$).}
\begin{equation}
\mathcal{H} = \frac12 \delta^{ab} \mathcal{B}_a^i \mathcal{B}_b^j \delta_{ij}
\end{equation}
Here, $(A^a_i) \equiv (A_i, Z_i)$ and
\begin{equation}
\mathcal{B}^{ i} _a=   \epsilon_{ab}\epsilon^{ijk} \partial_j A^b_k \, 
\end{equation}
are the ``magnetic fields'' ($\mathcal{B}^{ i} _1$ is actually the
standard electric field, while $\mathcal{B}^{i}_2$ is the standard
magnetic field \cite{Deser:1976iy,Deser:1997mz} -- up to signs that
depend on conventions).

The presymplectic $2$-form following from the action,
\begin{equation}
\Omega =\frac12 \int d^3x \, \epsilon_{ab} \epsilon^{ijk} \partial_i d_V A^a_j  \wedge d_V A^b_k \label{eq:PresymplecticInv}
\end{equation}
does not take the standard ``$d_Vp \wedge d_V q$'' canonical form. How
to deal with such $2$-forms is well known and recalled, for instance,
in appendix A of \cite{Henneaux:2018gfi}, of which we take the
notations.

The notable feature of $\Omega$ is that it is degenerate (hence the
``pre'' in ``presymplectic''). Indeed, vector fields of the following form  
\begin{equation}
\delta_X A^a_i = \partial_i \alpha^a, \qquad \alpha^a(\infty) = \textrm{constant}
\end{equation}
annihilate $\Omega$, 
\begin{equation}
i_X \Omega = 0.
\end{equation}
These correspond precisely to
proper gauge transformations (see Subsection \ref{subsec:Improper} below).

These are the only vector field with this property\footnote{In order to prove this statement, we assume that the potentials $A_i^a$ satisfy the asymptotic behaviour given in Subsection \ref{sec:boundary-conditions} below.}: setting $X = (\alpha^a_i)$, one finds
\begin{eqnarray} 
i_X \Omega &=& \frac12 \int d^3x \epsilon_{ab} \epsilon^{ijk} \partial_i \alpha^a_j   d_V A^b_k - \frac12 \int d^3x \epsilon_{ab} \epsilon^{ijk} \partial_i d_V A^a_j  \alpha^b_k  \nonumber \\
&=& \int d^3x \epsilon_{ab} \epsilon^{ijk} \partial_i \alpha^a_j   d_V A^b_k - \frac12 \oint_{S^\infty} d^2S_i \epsilon_{ab} \epsilon^{ijk} d_V A^a_j  \alpha^b_k  .\nonumber
\end{eqnarray}
One must solve the equation $i_X \Omega = 0$ for the $\alpha^a_i$'s.
Now, the $d_V A^b_k$ are independent $1$-forms, and so, the bulk term
vanishes if and only if the coefficient
$\partial_{[i} \alpha^a_{j]}$vanishes. This implies
$\alpha^a_i = \partial_i \alpha^a$. In order to abide by the asymptotic behaviour of $A_i^a$, the function $\alpha^a$ should tend to a $O(1)$-function that can depend on the angles as $r \rightarrow \infty$ , i.e., $\alpha^a = \xbar \alpha^a (\theta, \varphi) + O(r^{-1})$ for some function $\xbar \alpha^a$ on the $2$-sphere. The surface term can then be
rewritten as
$- (1/2) \oint_{S^\infty} d^2 S_i \epsilon_{ab} \epsilon^{ijk} d_V
(\partial_j A^a_k) \alpha^b$, which vanishes only if the $\alpha^b$'s
tend to a constant at infinity, i.e., if the functions $\xbar \alpha^a$ on the $2$-sphere reduce to their $0$-th spherical harmonic, $\xbar \alpha^a (\theta, \varphi) = \xbar \alpha^a_0$.

The zero vector fields of $\Omega$ (i.e., the phase space vector
fields that annihilate $\Omega$, $i_X \Omega = 0$) are thus precisely the vector
fields generating proper gauge transformations (Subsection \ref{subsec:Improper}).

\subsection{Hamiltonian vector fields}

One says that a phase space vector field $X$ (representing an
infinitesimal transformation through $\delta z^\alpha = X^\alpha$,
where $z^\alpha$ are the phase space variables) is Hamiltonian (and
that the transformation is canonical) if there exists a phase space
function $F$ such that
\begin{equation}
i_X \Omega = - d_V F \, .
\end{equation}
The function $F$ is called the generator of the transformation. This
is equivalent to the condition $\mathcal{L}_X \Omega = 0$, i.e., the
transformation generated by $X$ preserves the presymplectic form (we
assume trivial topology, more precisely, that every closed $2$-form is
exact).

Given a Hamiltonian vector field, the function $F$ is, as usual,
determined up to a constant. Because $\Omega$ is degenerate, however,
two new features arise. First, a function $F$ can be associated with a
Hamiltonian vector field only if
$i_{X_{a}} d_V F = \mathcal{L}_{X_a} F = 0$ for the zero vector fields
$X_a$ of $\Omega$, i.e., if it is invariant under the flow generated
by $X_a$. In our case, this means that $F$ has to be gauge invariant
(under proper gauge transformations). Second, given a function $F$
fulfilling this condition, the corresponding $X$ is determined up to a
combination of the $X_a$'s. In our case, this means that the
transformation generated by $F$ is determined up to a proper gauge
transformation. In particular, the zero phase space function
($F \equiv 0$) is associated with the Hamiltonian vector fields
defining the proper gauge transformations.

These properties are physically sensible. It turns out sometimes,
however, to be more convenient to deal with a true symplectic form,
i.e., an invertible $\Omega$. There are different ways to get one. One
way, which preserves spacetime locality, enlarges the phase space and
is described below.

\subsection{Boundary conditions}
\label{sec:boundary-conditions}

Before proceeding further, we shall specify the boundary conditions on
the fields. These are adapted from
\cite{Henneaux:2018gfi,Henneaux:2018hdj} and read:
\begin{equation}
A^a_i = \frac{\xbar A^a_i(\mathbf{n})}{r} + O\left(\frac{1}{r^2}\right)
\end{equation}
where $\mathbf{n}$ is the unit vector to the radial spheres and stands
therefore for coordinates on the unit sphere. Instead of standard
polar coordinates $(\theta, \varphi)$ which behave as
$\theta \rightarrow \pi - \theta$ and
$\varphi \rightarrow \varphi + \pi$ under the antipodal map, we shall
find it convenient to use coordinates $x^A$ which transform instead as
$x^A \rightarrow - x^A$. (Of course, neither $(\theta, \phi)$ nor
$(x^A)$ provide a single global chart.)

We impose in addition the following ``twisted parity condition'' on
the leading term of the vector potentials,
\begin{equation}
A^a_i (r, -\mathbf{n}) = A^a_i (r, \mathbf{n}) +\partial_i \lambda^a + O\left(\frac{1}{r^2}\right)
\end{equation}
for some $\lambda^a (\mathbf{n})$ that may assumed to be even. So, the
leading terms are even up to a gradient.  Differently put, the even part of $A^a_i $ is unrestricted, but the odd part must be a gradient to leading order.

This implies that the fields
are odd to leading orders,
\begin{equation}
\mathcal{B}^i_a = \frac{\xbar {\mathcal B}^i_a(\mathbf{n})}{r^2} + O\left(\frac{1}{r^3}\right), \qquad \xbar {\mathcal B}^i_a(-\mathbf{n}) = -\xbar {\mathcal B}^i_a(\mathbf{n})
\end{equation}
In spherical coordinates, the twisted parity conditions read
\begin{equation}
A^a_r = \frac{\xbar A^a_r(x^A)}{r} + O\left(\frac{1}{r^2}\right), \qquad A^a_A = \xbar A^a_A(x^B) + O\left(\frac{1}{r}\right)
\end{equation}
with 
\begin{equation}
\xbar A^a_r(-x^A) = - \xbar A^a_r(x^A), \qquad \xbar A^a_A(-x^B) = \xbar A^a_A(x^B) + \partial_A \lambda^{a} \, .
\end{equation}

With these boundary and parity conditions the presymplectic term
in the action is finite.  The logarithmic divergence, potentially present without parity conditions, is actually absent even if one allows a gradient in $A_i^a$ because the coefficient of $\dot{A}_i^a$ in the kinetic term is identically transverse.

\subsection{Improper gauge symmetries}
\label{subsec:Improper}

The boundary conditions  are invariant under gauge transformations 
\begin{equation}
\label{eq:GaugeTransfo}
\delta_\varepsilon A^a_i = \partial_i \varepsilon^a, \qquad \varepsilon^a = \xbar \varepsilon^a (x^A) + O\left( \frac{1}{r} \right),
\end{equation}
where there is no parity condition on $\xbar \varepsilon^a$. When
$\xbar \varepsilon^a (x^A) \not= 0$, the gauge transformation is
generically improper and defines a non trivial symmetry
\cite{Benguria:1976in}. More precisely, the generator
$G[\varepsilon^a]$, found through the equation $i_X \Omega = - d_V G$,
is given by
\begin{equation}
G[\varepsilon^a] = -\frac12 \oint_{S^\infty} d^2 S_i \,  \xbar{\mathcal{B}}^{ i}_a  \xbar \varepsilon^a
\end{equation}
as the above computation indicates. Since both $d^2 S_i = n_i d^2 S$ and $\xbar{\mathcal{B}}^{i}_a$
are odd, only the even part of the gauge parameters $\varepsilon^a$
contributes to the charges,
\begin{equation}
  \label{eq:Charges1}
  G[\varepsilon^a] = -\frac12 \oint_{S^\infty} d^2 S_i \,  \xbar{\mathcal{B}}^{ i}_a  \, \xbar \varepsilon^a_{\textrm{even}}  \, .
\end{equation}
Furthermore, because we consider source-free electromagnetism, the electromagnetic fluxes corresponding to constant $\varepsilon^a$'s in (\ref{eq:Charges1}) are zero as can
be seen by converting the surface integral to a volume integral
through Gauss formula.  This is in agreement with the observation
we made in section \ref{sec:action} that gauge parameters $\varepsilon^a$ that tend to a 
constant at infinity generate proper gauge transformations. The other angle-dependent $u(1)$
transformations have generically non-vanishing charges, however.

The improper gauge symmetries are abelian and form the infinite
dimensional algebra
$u(1)^{\textrm{even}} \oplus u(1)^{\textrm{even}}$. By the notation
$u(1)^{\textrm{even}}$ we mean the infinite-dimensional algebra of
gauge transformations parametrized by arbitrary even functions on the
$2$-sphere (and with no zero mode). We shall discuss below how the odd functions on the sphere
enlarge further the algebra.

It is worth stressing that the generator of improper gauge symmetries
is a pure surface term, without bulk piece. This is because, with the
degenerate presymplectic structure, the proper gauge transformations
(which can be viewed as the bulk part of the improper ones) have
through the equation $i_X \Omega = -d_V G$ a ``generator'' $G$ that
identically vanishes. The surface term by itself is a well defined
generator.

Another interesting feature that is exhibited by formula
(\ref{eq:Charges1}) is that both the electric and magnetic fluxes come
with a factor $1/2$. In hindsight this was to be expected if one
recalls that in the manifestly duality invariant formulation, both
electric and magnetic sources are minimally coupled to their
respective electromagnetic potentials, but only with half of the
strength of their charges (see \cite{Deser:1997mz} and Section \ref{sec:Sources} below). Gauss law for the
electromagnetic fields yield therefore half of the fluxes. Electric
and magnetic sources have also Dirac string type couplings, again with
half of the strength of the charges. The Dirac strings carry the other
half of the fluxes.

Finally, we note that the presymplectic form
(\ref{eq:PresymplecticInv}) evidently belongs to a family of presymplectic forms
that differ by boundary terms and yield the same equations of motion,
\begin{equation}
\Omega_\sigma =\frac12(1+\sigma)  \int d^3x  \epsilon^{ijk} \partial_i d_V A^1_j  \wedge d_V A^2_k - \frac12(1-\sigma)  \int d^3x  \epsilon^{ijk} \partial_i d_V A^2_j  \wedge d_V A^1_k\, . \label{eq:PresymplecticNonInv}
\end{equation}
with $\sigma \in [-1,1]$. The value $\sigma = 0$, considered in \cite{Deser:1997mz}, reproduces the
duality invariant presymplectic form considered here, while the extreme values
$\sigma = -1$ and $\sigma = 1$ correspond to the electric or magnetic
formulations, respectively (e.g., $\sigma = -1$ yields
$\Omega_{-1} = -\int d^3x \epsilon^{ijk} \partial_i d_V A^2_j \wedge
d_V A^1_k$ which is the standard presymplectic form
$\int d^3x d_V\pi^k \wedge d_V A_k$ of the usual electric formulation
with the electric potential $A_k \equiv A^1_k$, if one recalls that
the electric field $E^k = \epsilon^{ijk} \partial_i Z_j $ is minus the
momentum conjugate to $A_k$ ($A^2_k \equiv Z_k$ in the notations of
\cite{Deser:1976iy})).

Even though the difference between the presymplectic structures $\Omega_\sigma$ is
a mere surface term, these are physically inequivalent because the surface term in question does not vanish.  This leads to two important distinct features.   First the forms of the electric and magnetic couplings are different.  Both couplings are allowed, but they must be included differently dependong on the value of $\sigma$. An electric source minimally couples with strength $\frac12(1-\sigma)$ to the electric vector potential, the remaining of the coupling ($\frac12(1+\sigma)$) being accounted for by Dirac string type terms \cite{Deser:1997mz}.  Similarly, a magnetic source minimally couples with strength $\frac12(1+ \sigma)$ to the magnetic vector potential, the remaining of the coupling ($\frac12(1-\sigma)$) being accounted for by Dirac string type terms.  The symmetric case ($\sigma = 0$) has both types of sources and of couplings on the same footing.

Second,  the respective weights of the physically relevant improper gauge transformations are different. For all values of $\sigma \in (-1,1)$, the
improper gauge symmetries form the algebra
$u(1)^{\textrm{even}} \oplus u(1)^{\textrm{even}}$, and the generators
take the above form but with weights $\frac12 (1+\sigma)$ and
$\frac12 (1-\sigma)$, respectively. In the limiting cases
$\sigma = \pm 1$, one of the two $u(1)^{\textrm{even}}$'s becomes
proper because the corresponding generators vanish for all
configurations. The improper gauge transformations reduce to a single
algebra of angle-dependent (even) $u(1)$ transformations. So, for
$\sigma = -1$, the magnetic gauge transformations are all proper. In
that case, the electric flux is entirely carried by the electric field
(there is no Dirac string for electric sources) while the magnetic
flux of magnetic monopoles is entirely carried by the Dirac string.  Conversely,  for
$\sigma = +1$, the electric gauge transformations are all proper and the coupling of electric sources is entirely of Dirac string type.

\subsection{Equations of motion}

Stationarity of the action (\ref{eq:action}), $\delta S = 0$,  implies
\be
0 = \int dt \int d^3 x \epsilon_{ab} \epsilon^{ijk}\partial_i  \left(\dot{A}^b_k + \mathcal{B}^m_ c \delta_{mk} \delta^{cb}\right) \delta A^a_j + \frac12 \int dt \oint_{S_\infty} d^2 S_i \epsilon_{ab} \epsilon^{ijk} \dot{A}^b_k \, \delta A^a_j 
\ee
where we have dropped terms at the initial and final time boundaries and where we have used the boundary conditions as $r \rightarrow \infty$  to infer that $\mathcal{B}^m_ c \delta A^a_j \sim O(1/r^3)$ decays too fast to contribute to the surface integral at spatial infinity. Dropping boundary terms at $t = t_i$ and $t=t_f$ is legitimate provided one includes the appropriate surface terms there, along the lines discussed for instance in  \cite{Henneaux:1987hz}.  We will not dwelve on this well understood issue here, since we want to focus on the difficulties raised by the behaviour of the fields at spatial infinity.

The vanishing of the bulk term in $\delta S$ implies
\be 
\epsilon^{ijk}\partial_i  \left(\dot{A}^b_k + \mathcal{B}^m_ c \delta_{mk} \delta^{cb} \right) = 0 \qquad \Rightarrow \qquad \dot{A}^b_k =  -\mathcal{B}^m_ c \delta_{mk} \delta^{cb} + \partial_k A_0^b \label{Eq:EvolA}
\ee
for some arbitrary functions $A_0^b$ which are only restricted at this stage to be such that  $\partial_k A_0$ is of order $O\left(\frac{1}{r} \right)$, in order to preserve in time the asymptotic behaviour of $A^b_k$.

The vanishing of the surface term in $\delta S$ puts constraints on the leading term of the coefficient $\epsilon^{ijk} n_i \dot{A}^b_k \sim \epsilon^{ijk} n_i \partial_k A_0^b$ of  $\delta A^a_j$ in the surface integral.   Indeed, since the leading even part of the variation $\delta A^a_j$ is arbitrary, the leading even part of its coefficient should be equal to zero, or equivalently, the leading odd part of  $\partial_k A_0^b$ (coming from the leading even part of $A_0)$) must vanish.  This implies that the ambiguity in the time evolution, captured by the $A_0$-term in (\ref{Eq:EvolA}), is a {\em a proper gauge transformation}, as it should. There is no improper gauge transformation involved in the time evolution ambiguity once the Hamiltonian is given.  Without loss of generality, one may asymptotically fix the gauge and assume that there is no $O(1)$-piece in $A_0$,
\be
A_0 = \frac{\xbar A_0}{r} + O\left(\frac{1}{r^2}\right) \label{eq:AsA0}
\ee
It is easy to verify that there is no additional constraint coming from the leading odd part of the variation $\delta A^a_j$ so that the equations of motion are completely equivalent to (\ref{Eq:EvolA}) and (\ref{eq:AsA0}) (under the above partial gauge fixing).

An interesting consequence of the equations of motion is
\be
\dot{\xbar A}^a_r = 0
\ee
as one can see by expanding (\ref{Eq:EvolA}) with $k=r$ in powers of $r^{-1}$.

\section{Poincar\'e invariance}
\label{sec:Poincare}

\subsection{Boundary degrees of freedom}

In order to implement Poincaré invariance, one needs to introduce a
surface degree of freedom at infinity, $\xbar \Psi$, which is
conjugate to the gauge-invariant asymptotic value $\xbar A_r$ of the
radial component of the electric vector potential. This was explained
in \cite{Henneaux:2018gfi}, where it was also shown that this new
degree of freedom can be interpreted as the $O(1/r)$-term in the
asymptotic expansion of the temporal component $A_0$. In the
duality-symmetric formulation, one needs a $so(2)$ electric-magnetic
doublet $\xbar \Psi^a$.

Following \cite{Henneaux:2018gfi}, we thus add to the action the term
\begin{equation}
- \frac12 \int dt \oint d^2x \sqrt{\xbar \gamma} \, \delta_{ab} \, \xbar A^a_r \, \partial_t \xbar \Psi^b
\end{equation}
leading to
\begin{equation}
S[A^a, \xbar \Psi^a] = \int dt \left[\int d^3x \left(\frac12 \epsilon_{ab} \epsilon^{ijk} \partial_i A^a_j \dot{A}^b_k - \mathcal{H} \right) - \frac12  \oint d^2x \sqrt{\xbar \gamma} \, \delta_{ab} \, \xbar A^a_r \, \partial_t \xbar \Psi^b \right]
\end{equation}
The new symplectic structure is thus
\begin{equation}
    \Omega^{full}_d 
    =\frac12  \int d^3x \,  \epsilon_{ab} \epsilon^{ijk} \partial_i d_V A^a_j \, d_V A^b_k 
    - \frac 1 2 \oint d^2x \sqrt {\xbar \gamma }\, 
      \delta_{ab} d_V \xbar A^a_r d_V \xbar \Psi^b .
\end{equation}
The factor of one half present in the new boundary term matches the
factor of one half in the symplectic form $\Omega$, which is itself a
consequence of the boundary term difference between the usual
``electric'' symplectic structure and the duality invariant one that
we have pointed out.

Because $\xbar A^a_r$ is odd, only the odd part of $\xbar \Psi^a$
appears in the action. The even part of $\xbar \Psi^a$ is pure gauge.  One may either 
fix that gauge and assume e.g. that it is zero, so that
$
\xbar \Psi^a (-x^A) = - \xbar \Psi^a(x^A),
$
or one can chose not to fix that gauge and keep the even part of $\xbar \Psi^a$ arbitrary.   It turns out that this second approach is more convenient.  
Thus we have
\be
\xbar \Psi^a = \left(\xbar \Psi^a\right)^{\rm even} + \left(\xbar \Psi^a\right)^{\rm odd}, \qquad \left(\xbar \Psi^a\right)^{\rm even} = \hbox{ pure gauge}
\ee
where the odd part of $\xbar \Psi^a$ corresponds to physical
degrees of freedom.
Given the extra degeneracy of the extended symplectic form, a Hamiltonian function must be invariant under the corresponding gauge symmetry, i.e., be independent of $\left(\xbar \Psi^a\right)^{\rm even}$. 

The new degree of freedom and the new term in the action bring in additional equations, which are
\be
\dot{\xbar A}^a_r = 0, \qquad \left(\dot{\xbar \Psi}^a_r \right)^{\rm odd}= 0.
\ee
The first one follows from extremization with respect to $\xbar \Psi^a$ and is a consequence of the other equations of motion.  The second one follows from extremization with respect to $A_r^a$. 

\subsection{More improper gauge transformations}

The action is also invariant under arbitrary (time-independent) shifts of
$\xbar \Psi^a$  by an odd function,
\begin{equation}
\xbar \Psi^a \quad \rightarrow \quad \xbar \Psi^a + \xbar \mu^a, \qquad \xbar \mu^a(-x^A) = - \xbar \mu^a(x^A).
\end{equation}
These  are canonical transformations with generators
\begin{equation}
G_\mu = - \frac12 \oint d^{2}x \sqrt{\xbar \gamma} \delta_{ab} \xbar \mu^a \xbar A^b_r  . \label{eq:Charges2}
\end{equation}
These transformations are improper gauge transformations
since their generators generically do not vanish.

The total set of soft charges is thus given by
\begin{equation}
  G[\xbar \varepsilon^a, \xbar \mu^a] = - \frac12 \oint_{S^\infty} d^2 S_i \,  \xbar{\mathcal{B}}^{ i}_a  \, \xbar \varepsilon^a
  - \frac12  \oint d^{2}x \sqrt{\xbar \gamma} \delta_{ab} \xbar \mu^a \xbar A^b_r\label{eq:Charges3}
\end{equation}
with $\xbar \varepsilon^a= \xbar \varepsilon^a_{\textrm{even}}$ and $\xbar \mu^a = \xbar \mu^a_{\textrm{odd}}$.
The generators reduce to surface integrals at infinity, without bulk
term, because Gauss' law has been solved for and so is identically
satisfied. As we explained, this does not prevent the surface
integrals to be well defined canonical generators. In the pure
electric formulation where Gauss' law is not solved for, it is natural
not only to extend the boundary degree of freedom $\xbar \Psi^a$ in the bulk (as we shall actually do below), but also to introduce
its bulk conjugate momentum, which is constrained to vanish, preserving a symmetric treatment of the two gauge symmetries.  There
is no motivation for performing this second step  here, since there is no bulk constraint associated with the standard gauge invariance to begin with.

The improper gauge transformations commute and the algebra of their
charges does not acquire a central extension. This is is because the generators are invariant under both proper and improper gauge symmetries. It must be  contrasted with the approach of \cite{Hosseinzadeh:2018dkh,Freidel:2018fsk}. Note that the zero modes of the generators identically vanish (no source) and so a central charge extension mixing electric and magnetic generators of the form 
$$ [G[\xbar \varepsilon^a, \xbar \mu^a],  G[\xbar \varepsilon'^a, \xbar \mu'^a]] \sim  \epsilon_{ab} \int_{s^\infty} d^2x \varepsilon^a \varepsilon'^b$$
(say, or of similar undifferentiated type), is not possible since the right-hand side must identically vanish when $\varepsilon^a =$ constant.

\subsection{Poincar\'e transformations}

One can now easily verify
Poincar\'e invariance. The steps are the same as in the electric
formulation \cite{Henneaux:2018gfi}. We denote the normal and tangential components of the Poincar\'e transformations by $\xi$ and $\xi^{i}$, respectively.  One has 
\be \xi = rb(x^A) + T \ee 
with 
\be
D_A D_B b + \xbar \gamma_{AB} b = 0
\ee
 and $T =$ constant, as well as $\xi^i = {b^{i}}_j x^j + W^i$ with $b_{ij} = - b_{ji}$ and $W^i =$ constant.  Here, $D_A$ is the covariant derivative on the unit sphere with standard round metric $\xbar \gamma_{AB}$.   In spherical coordinates, the tangential components read
 \begin{equation}
\xi^A = Y^A + \frac 1 r \xbar \gamma^{AB} \d_B W, \quad
    \xi^r = W, 
  \ee
  where $Y^A$ and $W$ are functions on the sphere such that
  \be \cL_Y \xbar \gamma = 0, 
    \quad D_A D_B W + \xbar \gamma_{AB} W = 0
\end{equation}
(the $Y^A$'s define Killing vectors on the sphere).

For normal deformations we assume that the fields transform as
\begin{equation}
  \label{eq:normdef}
      \delta_{\xi, 0} A^{a}_i =  - \frac \xi {\sqrt g} \mathcal{B}^{a}_i  + \d_i(\xi \Psi^{a}) 
\end{equation}
(with $\mathcal{B}^{a}_i \equiv \mathcal{B}_{b}^j \delta^{ab} \delta_{ij} $) and
\begin{align}
 \delta_{\xi, 0} \left(\xbar\Psi^{a}\right)^{\rm odd} =   D^A\Big(b \left(\xbar A^{a}_A\right)^{\rm odd} \Big), \qquad     \delta_{\xi, 0} \left(\xbar\Psi^{a}\right)^{\rm even} = 2 b \xbar A^{a}_r + D^A\Big(b \left(\xbar A^{a}_A\right)^{\rm even} \Big) \, .
      \label{eq:normdef22}
\end{align}
The function
$\Psi^{a}$ appearing in (\ref{eq:normdef}) is an extension inside the bulk of the boundary variable $\xbar \Psi^a$, in the following sense,
\begin{equation}
  \Psi^{a} = \frac 1 r \xbar \Psi^{a} + O(r^{-2}) \, .
\end{equation}
This matching condition is the only condition on $\Psi^{a}$, since two different extensions will then differ by a proper gauge transformation. Note that for non zero boosts ($b \not=0$), the linear growth of $b$ compensates the $\frac1r$ decay of $\Psi^a$ so that the term  $\xi \Psi^{a}$ tends to the $O(1)$ even function $b \xbar \Psi^a$ at infinity and induces therefore a nontrivial improper gauge transformation in $\delta_{\xi, 0} A^{a}_i$.  

Similarly, the transformation (\ref{eq:normdef22}) involves both an improper gauge part, namely, $ \delta_{\xi, 0} \left(\xbar\Psi^{a}\right)^{\rm odd}$ and a proper gauge  one, namely $ \delta_{\xi, 0} \left(\xbar\Psi^{a}\right)^{\rm even}$.   That second one is completely arbitrary.  The gauge choice made in (\ref{eq:normdef22}) is convenient as it can be related to the Lorentz gauge (see below).

For convenience we also provide the leading contribution of \eqref{eq:normdef} in spherical coordinates
\begin{align}
      \delta_{\xi, 0} \xbar A^{a}_r = - b\, \epsilon^{a}_{\phantom ab} e^{AB} \d_A \xbar A^{b}_B \qquad
    \delta_{\xi, 0} \xbar A^{a}_A =   - b \,\epsilon^{a}_{\phantom ab}\gamma_{AB} e^{BC} \d_C \xbar A^{b}_r + \d_A(b \xbar \Psi^{a}),
\end{align}
with $\sqrt{\xbar \gamma} e^{AB} \equiv \epsilon^{rAB}$.

One can motivate the form of $ \delta_{\xi, 0} A^{a}_i$ as follows.  If we want to view $(A_\mu^a)$ as the components of $4$-vectors, their Poincar\'e transformations should coincide with their Lie derivatives on-shell and up to gauge transformations.  Now, with $(\xi^\mu) = (\xi, 0)$,
\be
{\cal L}_\xi A^a_k  = \xi \partial_0 A^a_k + \partial_k \xi A^a_0 =  \xi (-\mathcal{B}^{a}_k + \d_k A_0^{a}) +  \partial_k \xi A^a_0 = -\xi \mathcal{B}^{a}_k + \d_k (\xi A_0^{a}),
\ee
upon use of the equations of motion.  Thus, if we fix further the freedom in $A_0^a$ such that $\xbar A_0^a = \xbar \Psi^a$, which is permissible and in fact already considered in our earlier work \cite{Henneaux:2018gfi,Henneaux:2018hdj}, one finds 
\be
{\cal L}_\xi A^a_k  = -\xi \mathcal{B}^{a}_k + \d_k (\xi \Psi^{a})
\ee
which is precisely (\ref{eq:normdef}) (recall that $g=1$ in Cartesian coordinates).

When the condition $A_0^a = \Psi^a$ holds, the Lorenz gauge $\partial^\mu A_\mu = 0$ holds asymptotically, since $\partial^\mu A_\mu = \partial^0 A_0 + O(1/r^2) = \frac{\partial^0 \xbar \Psi^a}{r} + O(1/r^2) = O(1/r^2)$ on account of $\dot{\xbar \Psi}^a = 0$.  This provides a motivation for the transformation rule for $\xbar \Psi ^a$ -- in addition to the fact that the complete transformation must be canonically generated. Indeed, when the Lorentz gauge holds, the transformation of $A_0^a$ is
$
{\cal L}_\xi A^a_0  = \xi^\mu \partial_\mu A^a_0 + \partial_0 \xi^\mu A^a_\mu
$.
Now, with $\xi^\mu \frac{\partial}{\partial x^\mu}= x^k  \frac{\partial}{\partial x^0} + x^0  \frac{\partial}{\partial x^k}$ (boost along the $k$-th direction), one has $(\xi^\mu) = (x^k, 0)$ and $(\partial_0 \xi^\mu) = (0, \delta^m_k)$ on the slice $x^0 = 0$ (translating the slice in time will only generate additional translation terms that affect only the subleading terms of the fields), and so
$
{\cal L}_\xi A^a_0  = x^k \partial_0 A^a_0 +  A^a_k = x^k \partial^m A_m^a + A^a_k = \partial^m (x_k A_m^a)
$,
i.e., 
\be
{\cal L}_\xi A^a_0 = D^m(\xi A^a_m)
\ee
with $\xi \equiv x^k \equiv b r$ with $b$ the function of the angles corresponding to $x^k$.  Expanding this relation in polar coordinates and keeping the leading term, one gets exactly (\ref{eq:normdef22}).

The associated generator, which is invariant under proper gauge transformations and is therefore indeed an acceptable canonical generator, is given by
\be
    P^{full}_{\xi,0} = 
    \int d^3x \, \xi \, \mathcal{H}
    + \oint d^2x \, b \frac {\sqrt {\xbar \gamma}} 2 
      \left(  - \delta_{ab} (\xbar A^{a})^A \d_A \xbar A^{b}_r
      + \epsilon_{ab}  \, e^{AB} \d_A \xbar A^{a}_B \xbar \Psi^{b}
    \right) \, ,
\ee
Note that only the odd part of $\xbar \Psi^{a}$  appears in this expression as requested by gauge invariance, since $\sqrt{\xbar\gamma} \epsilon_{ab} e^{AB} \d_A \xbar A^{b}_B = \lim_{r\to \infty} \mathcal B^r_a$ is even.   

Because the (pre-)symplectic form is degenerate, the transformation generated by $ P^{full}_{\xi,0}$ is defined up to proper gauge transformations, which can be chosen as one pleases.  By contrast the improper gauge transformations entering the transformations is not arbitrary.  
In particular, the asymptotic term $\d_A(b \xbar \Psi^{a})$ in the transformation of $\xbar A^a_A$  is an improper gauge transformation which is determined by the generator.  Two different bulk extensions of that improper gauge transformation differ by a proper gauge transformation.  This the ambiguity in the bulk field $\Psi^a$.

For the spatial component $\xi^{i}$ we assume that $\xbar \Psi^{a}$
transforms as a scalar under rotations. We can then write the full
generator for the spatial translations and rotations as
\begin{multline}
  P^{full}_{0,\xi^i} =  \frac{1}{2}\int d^3x \,
  \epsilon_{ijk} \epsilon^{ab} \xi^i \mathcal{B}^{j}_{a}\mathcal{B}^{k}_{b} 
    + \oint d^2x \, Y^A \frac {\sqrt\gamma} 2
    (
  -\epsilon_{ab} \xbar A^{a}_A \,  e^{BC} \d_B \xbar A^{b}_C  
    +\delta_{ab} \xbar \Psi^{a} \, \d_A \xbar A^{b}_r
    ),
\end{multline}
which is again easily seen to be invariant under proper gauge transformations.

With these generators at hand one can show that the system is indeed
invariant under Poincaré transformations and fulfill the Poincaré
algebra.

\subsection{Poincaré transformations of the improper gauge generators}

The generators of the Poincaré and gauge transformations span a
semidirect sum, with the gauge transformations being an abelian ideal.
The action of the Lorentz algebra controlling the semi-direct sum is
given by
\begin{align}
    \delta_{\xi,\xi^{i}} \xbar \epsilon^{a} = 
  Y^A\d_A \xbar \epsilon^{a} + b \xbar \mu^{a}, \qquad
      \delta_{\xi,\xi^{i}} \xbar \mu^{a} = 
      Y^A \d_A \xbar \mu^{a} + D^A(b \d_A \xbar \epsilon^{a}) \, .
\end{align}
At this point we have two pairs of functions of definite parity, the
odd $\xbar \mu^{a}$ and the even $\xbar \epsilon^{a}$. In order to compare  our results with the null infinity analysis, it remains to
argue that these functions combine to form  functions on the sphere with
no definite parity. To that end, one can adapt \cite{Troessaert:2017jcm} or Appendix C
in~\cite{Henneaux:2018gfi}, where this was explicitly shown for the
``common'' formulation of electrodynamics and which generalizes to the
case at hand.

\subsection{$SO(2)$ duality generator}
\label{sec:so2-dual-gener}

One key feature of the double potential formulation of
electrodynamics is the manifest $SO(2)$ duality invariance that
acts locally on the canonical variables~\cite{Deser:1976iy}.  This duality transformation also rotates the asymptotic fields $\xbar \Psi^a$.
Explicitly, the rotation
\begin{align}
  \begin{pmatrix}
    A^{\prime 1} \\
    A^{\prime 2}
  \end{pmatrix}
  =
  \begin{pmatrix}
    \cos \phi & \sin \phi \\
    -\sin \phi & \cos \phi
  \end{pmatrix}
                 \begin{pmatrix}
                   A^{1} \\
                   A^{2}
                 \end{pmatrix}, \qquad
 \begin{pmatrix}
    \xbar \psi^{\prime 1} \\
    \xbar \psi^{\prime 2}
  \end{pmatrix}
  =
  \begin{pmatrix}
    \cos \phi & \sin \phi \\
    -\sin \phi & \cos \phi
  \end{pmatrix}
                 \begin{pmatrix}
                    \xbar \psi^{1} \\
                    \xbar \psi^{2}
                 \end{pmatrix}
\end{align}
leaves the action invariant.  Through the formula $i_X \Omega = - d_V Q$, one finds that it is generated by
\begin{align}
  Q= - \frac{1}{2} \int d^{3}x \delta_{ab} \epsilon^{ijk}  A^a_i \partial_j A^b_k - \frac12 \oint d^2x \sqrt{\xbar \gamma} \epsilon_{ab} \xbar A^a_r \xbar \Psi^b \, .
\end{align}
The first term is the Chern--Simons term of \cite{Deser:1976iy}.  The second appears beause of the extra surface degrees of freedom $\xbar \Psi^a$.

Without the parity conditions of Section \ref{sec:boundary-conditions}
this expression would be logarithmically divergent. However, since we
impose that the leading term in $A^{a}_{i}$ is even up to a gradient (which contributes a manifestly finite surface term since the magnetic field is identically transverse), the coefficient of the logarithmically divergent term is equal to zero and 
we get a well defined symmetry generator.

The algebra of the $so(2)$-duality generator with the generators of the improper gauge transformations can easily be worked out,
\be
[G[\xbar \varepsilon^a , \xbar \mu^a], Q] = G[\xbar \varepsilon^b {\epsilon_b}^a, \xbar \mu^b{\epsilon_b}^a].
\ee

\section{Sources}
\label{sec:Sources}

The inclusion of sources follows the pattern of \cite{Deser:1997mz}.   In the duality-symmetric formulation, half of the coupling is of standard minimal type, while the other half follows the Dirac procedure and involves Dirac strings.  To comply with the asymptotic parity conditions, these strings will be chosen symmetrically in the asymptotic region, i.e., half of the flux carried by the string (a quarter of the total flux for that matter) will be brought from one direction, while the other half (quarter) will be brought from the antipodal direction (asymptotically).  This is the same set up as for gravity \cite{Bunster:2006rt}.

The detailed form of the action with sources included is given in \cite{Deser:1997mz}.  What replaces the identity 
\be
\partial_i {\mathcal B}^i_a \equiv 0
\ee
is the identity with sources
\be
\partial_i {\mathcal B}^i_a (\vec{x})\equiv \sum_A q_a^{(A)} \delta^{(3)}(\vec{x} - \vec{z}^{(A)})
\ee
where $A$ runs over the various sources, located at $\vec{z}^{(A)}$.  It follows that
\be
\oint_{S^\infty} d^2 S_i  {\mathcal B}^i_a = \sum_A q_a^{(A)}.
\ee
is a ``c-number'', having zero bracket with anything. 

The action is invariant under the same set of gauge transformations $\delta A^a_i = \partial_i \varepsilon^a$, which must be supplemented by the phase transformation $\delta p^{(A)}_i = - q_a^{(A)} \frac{\partial \varepsilon^a}{\partial z^{i (A)}}$ for the momenta conjugate to the position of the charged particles.  The generators of the gauge transformations take the same form as before, with improper gauge transformations characterized by non-vanishing surface integrals at infinity.  Because the zero mode of the magnetic field is a $c$-number,  there is again no central charge in the abelian algebra of the improper gauge generators, which is unchanged.

In addition to the gauge transformations, the theory is also invariant under arbitrary shifts of the Dirac strings (provided they remain attached to the sources). Non vanishing displacement of the Dirac strings at infinity is a pure gauge transformation, with zero generator.  It is useful not to fix this gauge and to allow non zero displacements at infinity, as this enables one to control more easily Poincar\'e invariance.  Indeed, under Lorentz boosts, the strings will naturally change their orientation and it is better not to impose a condition that it should always run, say, along the $z$-axis.

We close this section by noting that a different way to include sources was developed in \cite{Barnich:2007uu}.  It would be of interest to define consistent asymptotic conditions in that approach too.

\section{Conclusions}
\label{sec:Conclusions}

One of the main motivations for studying the asymptotic behaviour of the fields at spatial infinity, on Cauchy hypersurfaces, is to get better tools to understand the structure of the space of physical states, a tricky issue in gauge theories. States are indeed naturally defined on Cauchy hypersurfaces.  Physical states involves dressings \cite{Dirac:1955uv}.  Both electric and magnetic dressings are necessary in the presence of magnetic poles \cite{Antunovic:1984td}.  Most dressings are usually chosen not to fulfill any particular parity conditions \cite{Giddings:2019ofz,Choi:2019sjs}, leading to divergences in the boost generators, which must be regulated. Given the ambiguity in the dressing of physical states, one might consider dressings that fulfill our (twisted) parity conditions.  This has the advantage of avoiding divergences in Lorentz boosts altogether, as well as logarithmic behaviour at null infinity of some components of the fields which might lead to problems \cite{Henneaux:2018gfi,Henneaux:2018hdj}.  There is to our knowledge no indication that imposing such parity conditions is a physical limitation. 

\vspace{.3cm}

\noindent
Note added: after this paper was completed, we became aware of the interesting paper \cite{Bhattacharyya:2017obx} where the duality-symmetric formulation is also considered, but at null infinity, with almost no overlap with our construction.

\section*{Acknowledgments} 

We thank Glenn Barnich, Oscar Fuentealba, Victor Lekeu, Sucheta Majumdar, Javier
Matulich, Stefan Prohazka and Friedrich Sch\"aller for insightful discussions.

The research of MH is supported by the ERC Advanced Grant
``High-Spin-Grav'' and by FNRS-Belgium (Convention FRFC PDR T.1025.14
and Convention IISN 4.4503.15). During part of this work, CT was supported by the Max Planck Institute for Gravitational Physics (Albert Einstein Institute) in Potsdam.

\bibliographystyle{utphys}

\end{document}